\documentclass[10pt,twocolumn]{article}
\usepackage{graphicx} 
\usepackage{authblk}
\usepackage{float}
\usepackage{braket}
\usepackage{mathtools}
\usepackage{array}
\usepackage[acronym]{glossaries}

\newtheorem{theorem}{Theorem}
\usepackage{amsmath}
\usepackage{mathrsfs}
\usepackage{xcolor}
\usepackage{amssymb}
\usepackage{verbatim}
\usepackage{mathabx}
\usepackage[font=small,labelfont=bf]{caption}
\usepackage[super]{natbib}
\usepackage[hidelinks]{hyperref}
\usepackage[normalem]{ulem}
\usepackage[titletoc,toc,title]{appendix}

\makeglossaries

\title
{Quantum Information reveals that orbital-wise correlation is essentially classical in Natural Orbitals}

\author[1,2]{Davide Materia}
\author[1,2] {Leonardo Ratini}
\author[3]{Celestino Angeli}
\author[1]{Leonardo Guidoni
\thanks{leonardo.guidoni@univaq.it}}

\affil[1]{
Dipartimento di Scienze Fisiche e Chimiche, Universit\`a degli Studi dell’Aquila, Coppito, L’Aquila, Italy }

\affil[2]{Dipartimento di Ingegneria e Scienze dell'Informazione e Matematica\\ Universit\`a degli Studi dell'Aquila, Coppito, L'Aquila, Italy}

\affil[3]{Dipartimento di Scienze Chimiche, Farmaceutiche ed Agrarie, Università degli Studi di Ferrara, Italy}

\date{\today}

\begin{document}

\newacronym{dmrg}{DMRG}{Density Matrix Renormalization Group}
\newacronym{dm}{DM}{Density Matrix}
\newacronym{hf}{HF}{Hartree-Fock}
\newacronym{mp2}{MP2}{Møller-Plesset perturbation theory of second order}
\newacronym{ci}{CI}{Configuration Interaction}
\newacronym{mrci}{MRCI}{Multi-Reference Configuration Interaction}
\newacronym{cisd}{CISD}{CI with singles and doubles}
\newacronym{cc}{CC}{Coupled Cluster}
\newacronym{ccsd}{CCSD}{CC with singles and doubles}
\newacronym{scf}{SCF}{Self Consistent Fields}
\newacronym{sd}{SD}{Slater Determinants}
\newacronym{mo}{MO}{Molecular Orbitals}
\newacronym{so}{SO}{Spin Orbitals}
\newacronym{hfco}{HFCO}{Hartree-Fock Canonical Orbitals}
\newacronym{no}{NO}{Natural Orbitals}
\newacronym{ino}{INO}{Iterative Natural Orbitals}
\newacronym{tpcpm}{TPCPM}{Trace Preserving Completely Positive Map}
\newacronym{cas}{CAS}{Complete Active Space}
\newacronym{mrpt2}{MRPT2}{Multi-Reference Perturbation Theory}
\newacronym{fci}{FCI}{Full CI}

\twocolumn[
\begin{@twocolumnfalse}
\maketitle
\begin{abstract}

The intersection of Quantum Chemistry and Quantum Computing has led to significant advancements in understanding the potential of using quantum devices for the efficient calculation of molecular energies. Simultaneously, this intersection is enhancing the comprehension of quantum chemical properties through the use of quantum computing and quantum information tools.

This paper tackles a key question in this relationship: Is the nature of the orbital-wise electron correlations in wavefunctions of realistic prototypical cases classical or quantum?
We delve into this inquiry with a comprehensive examination of molecular wavefunctions using Shannon and von Neumann entropies, alongside classical and quantum information theory. Our analysis reveals a notable distinction between classical and quantum mutual information in molecular systems when analyzed with Hartree-Fock canonical orbitals. However, this difference decreases dramatically, by approximately 100-fold, when Natural Orbitals are used as reference. 

This finding suggests that wavefunction correlations, when viewed through the appropriate orbital basis, are predominantly classical. This insight indicates that computational tasks in quantum chemistry could be significantly simplified by employing Natural Orbitals. Consequently, our study underscores the importance of using Natural Orbitals to accurately assess molecular wavefunction correlations and to avoid their overestimation. In summary, our results suggest a promising path for computational simplification in quantum chemistry, advocating for the wider adoption of Natural Orbitals and raising questions about the actual computational complexity of the multi-body problem in quantum chemistry.

\end{abstract}
\vspace{1cm}
\end{@twocolumnfalse}

]

\section{\label{sec:intro}Introduction}
In recent years, there has been increasing interest in applying tools from quantum information theory to analyze chemical systems within the field of Quantum Chemistry~\cite{Hartree1928, Szabo2012, barcza2011, Ding2021, ding2023b}. This interest has led to a new understanding of known chemical properties~\cite{Boguslawski2014, Ding2020} and the development of efficient computational tools that focus computational efforts where they are most needed~\cite{Stein2016, Bensberg2023, Ding2023}.

Part of this growing interest is likely caused by the recent encounter of quantum chemistry with quantum computing~\cite{physchem}, a rapidly growing field that promises to bring great computational advantages to all computational sciences \cite{Feynman1982,Shor1997,Banuls2020,Hussain2020, Biamonte2017, Bordoni2023} and in particular to quantum chemistry \cite{Aspuru-Guzik2005, Grimsley2019, Ratini2022, Egger2023}.
Such an encounter, in addition to the direct application of quantum computing to Quantum Chemistry, promotes an exchange of ideas and technical tools in both directions, strengthening the synergy between quantum and classical computing systems..

This exchange has predominantly involved the transfer of concepts from quantum chemistry to quantum computing~\cite{Benfenati2021}, such as, for instance, the Unitary Coupled Cluster (UCC) method, which adapts the concept of excitations from the Coupled Cluster method into a variational circuit suitable for quantum devices~\cite{Hoffmann1998, Kutzelnigg1991, Cooper2010, Evangelista2011, Whitfield2011, Barkoutsos2018, Romero2019}. The analysis proposed here moves in the opposite direction, using typical quantum computing tools to better understand quantum chemical properties.

In the context of near-term quantum computing devices, the wavefunction ans\"atze inspired by quantum chemistry and used in the Variational Quantum Eigensolver require performances that are still far out of reach. For this reason, alternative methods based on heuristic variational circtuits are preferred because more effective, albeit the physical meaning of the wavefunction is lost ~\cite{Kandala2017, rattew2020, Tang2021, Ratini2022, Materia2024}.
Regardless of the chosen ansatz, different studies have shown that the choice of the reference orbitals significantly impacts the computational efficiency~\cite{Sokolov2020a, Ratini2022}. 

Within this context, \acrfull{no}, a wavefunction-dependent orbital set that diagonalizes the one body reduced density matrix, have long played an important role related to the conjectured simplification in the expression of the wavefunction as hypothesized by L\"odwin in 1956 ~\cite{lowdin}. Recently, Natural Orbitals have been also revealed to be a winning choice in empirical circuits for Quantum Computing, since they are the orbitals that optimize the energies of these ansatzes~\cite{Ratini2023b, Materia2024}. Notwithstanding the amount of studies exploiting the \acrshort{no}, so far no direct numerical study has tackled the role played by this one-electron basis set on the quantum information properties.



With growing interest in quantum information theory, these analyses have been applied to understand entanglement and correlation in molecular systems\cite{Boguslawski2014, Ding2021}. Interestingly, most of the existing works in the field are based on a Matrix Product State approach with analyses performed on 
 \acrfull{dmrg}\cite{White1992, White93, Schollwock2005, SCHOLLWOCK2011, Stein2016}, calculations, whereas  all \acrfull{ci} methods are currently mostly ignored.

In the present study, thanks to the SparQ~\cite{materia2024c} method, we fill this gap for \acrshort{ci}-based method for quantum information analysis. We employ an analysis with orbital correlation on configuration interaction wavefunction using the von Neumann entropy and Mutual Information, prominent tools in quantum information theory. Anticipating our result, we demonstrate, both mathematically and numerically, that classical and quantum correlations are almost identical only in the case we use Natural Orbitals as reference for the wavefunctions.

The paper is structured as follows: subsection~\ref{sec:qit} introduces the entropy and mutual information, together with the necessary background on classical quantum information theory. We then delve into an explanation of the importance of Natural Orbitals in \ref{subsec:no}, followed by the core concept of this work detailed in~\ref{subsec:1-2dm}, which expands from previous works on the subject, quantum information theory and natural orbitals. Computational details are outlined in section~\ref{subsec:computationaldetails}, followed by results and discussion in section~\ref{sec:results}. 

\section{\label{sec:tb}Methods and Theoretical Background}

\subsection{\label{sec:qit}Quantum and classical Mutual information}

In quantum information theory, one usually works in a Hilbert space in which each vector of the space represents a wavefunction. 
In the context of quantum computing, this Hilbert space is composed of multiple \textit{separable} two-dimensional Hilbert spaces ($\mathscr{H}_i$) called \textit{qubits}. The whole space of $N$ qubits $\mathscr{H}_N$ is
\begin{equation}
    \mathscr{H}_N=\bigotimes_{i}^{N}\mathscr{H}_i
    \label{eq:tensordot}
\end{equation}
so that the dimension of $\mathscr{H}_N$ is $2^N$ \cite{Nielsen2010}.

Given the product structure of \eqref{eq:tensordot}, a wavefunction $\ket{\phi} \in \mathscr{H}_N$ can be written as:
\begin{equation}
    \label{eq:wf}
\begin{split}
    &\ket{\phi}=\sum_{i_0,\cdots i_j,\cdots i_{N-1}} k_{i_0,\cdots i_j,\cdots i_{N-1}}\\
    & \qquad \ket{i_0}\otimes\cdots \ket{i_j}\otimes\cdots \ket{i_{N-1}}
\end{split}
\end{equation}
Where each $\ket{i_j}$ spans a local computational basis. One can then also define the Density Matrix of the same state as:

\begin{equation}
\label{eq:dm}
    \begin{split}
    &\qquad\qquad\qquad\qquad\rho_{\ket{\phi}}=\ket{\phi}\bra{\phi}\\
    &= \sum_{i_0,\cdots i_j,\cdots i_{N-1}} \sum_{i_0',\cdots i_j',\cdots i_{N-1}'} k_{i_0,\cdots i_j,\cdots i_{N-1}} k^{*}_{i_0',\cdots i_j',\cdots i_{N-1}'}\\
    &\ket{i_0}\otimes\cdots \ket{i_j}\otimes\cdots \ket{i_{N-1}}
    \bra{i_0'}\otimes\cdots \bra{i_j'}\otimes\cdots \bra{i_{N-1}'}
    \end{split}
\end{equation}
This description of a state is used to account for any general quantum or statistical property of a state, but it comes with a quadratic cost price in terms of components of the wavefunction $\phi$. 
Technically, this density matrix lays into the subspace of \textit{Endomorphisms} of $\mathscr{H}_N$, but to avoid overloading the notation, in the following this space will also be identified with the symbol $\mathscr{H}_N$.
For a bipartite quantum system with Hilbert spaces $\mathscr{H}_A$ and $\mathscr{H}_B$, with $A,B$ subsets of qubits, the partial trace over subsystem $B$ over the density matrix $\rho_{AB}$, is defined as:

\begin{equation}    
\begin{split}
    &\rho_A = \text{tr}(\rho_{AB})_{B}=\sum_{i_{B_0}}\cdots\sum_{i_{B_{b-1}}} \\
    &\bra{i_{B_0}}\otimes\cdots\bra{i_{B_{b-1}}} \rho_{AB}\ket{i_{B_0}}\otimes\cdots\ket{i_{B_{b-1}}}
\end{split}
\label{eq:partial_trace}
\end{equation}

Where $\ket{i_{B_j}}$ are the computational basis states of the b qubits of the set B.

A short clarification is necessary at this point. The concept of \textit{reduced density matrix} has different meanings in quantum information theory and in quantum chemistry. In quantum information theory, the term is often used to define a partially traced matrix (following the definition reported above) in place of the name  \textit{traced density matrix} used in this work. However, the same name in quantum chemistry is used to define a different object, the 1-body reduced density matrix, defined over a $\ket{\psi}$ lying in the Fock space $\mathcal{F}$ of the $N$ fermionic modes.

\begin{equation}
    \rho_{\psi,kl}=\prescript{}{\mathcal{F}}{\bra{\psi}}a^{\dag}_{k}a_{l}\ket{\psi}_{\mathcal{F}}, \quad k,l \in [0,N-1]
    \label{eq:1brdm}
\end{equation}
 These two objects have been related in the past \cite{Rissler2006, Boguslawski2015}, and a further comparison is reported in Sec.~\ref{subsec:1-2dm}.  They both directly depend on the wavefunction, but they still are different objects by dimension, definition and trace normalization. 

In this work, the main use of the traced density matrix is to calculate the von Neumann (or quantum) mutual information.\cite{vonneumann1996}

\textbf{von Neumann (or quantum) mutual information $\mathbf{I_{vN}(A,B)}$.}
\textit{Consider two Hilbert spaces $\mathscr{H}_A$ and $\mathscr{H}_B$, and let $\rho_{AB}\in \mathscr{H}_A \otimes \mathscr{H}_B$ be a density matrix. The quantum mutual information $I$ of $\rho_{AB}$ is given by:}
\begin{equation}
    I_{vN}(A,B) = S_{vN}(A) + S_{vN}(B) - S_{vN}(A,B)
    \label{eq:qmi}
\end{equation}
where $S$ represents the von Neumann entropy defined as
\begin{equation}
\begin{split}
    S_{vN}(A,B) &= -\operatorname{tr}(\rho_{AB}\log(\rho_{AB}))\\
    S_{vN}(A) &= -\operatorname{tr}(\rho_{A}\log(\rho_{A}))\\
    \rho_{A} &= \operatorname{tr}_{B}(\rho_{AB})
\end{split}
\label{eq:entropy}
\end{equation}

Clearly, these equations contain a dependency on the state which is represented by $\rho$. In the following, unless specifically necessary as in Appendix \ref{app:A}, this dependency will not be underlined by having $\rho$ as an input, since the main variables for our discussion are the space variables $A$ and $B$.\\ 
From equation~\eqref{eq:entropy}, for an N-partite Hilbert space $\mathscr{H}_N = \bigotimes_i^N\mathscr{H}_i$ and a quantum state expressed by its $\rho \in \mathscr{H}_N$, the element of the mutual information matrix $I_{vN}(i,j)$ is obtained by tracing out all subsystems but $i,j$. In our use-case, the subsystems are $N$ spin-orbitals mapped on equally many qubits by means of a fermion-to-qubit mapping~\cite{Jordan1928}. Finally, we note that the matrix is symmetric, with the values along the diagonal being undefined.

The mutual information defined in eq.~\eqref{eq:qmi} has been used in the past \cite{Huang2005, Stein2016, Ding2021} for its flexibility in describing correlation and for its further ability to discern high levels of entanglement.

\textbf{Shannon (or classical) mutual information $\mathbf{I_{Sh}(A,B).}$}
In classical systems we are dealing with statistical combinations of states. In this case, correlations between two systems $A$ and $B$ can be investigated taking into account the Shannon (or Classical)\cite{Shannon1948} mutual information $I_{Sh}(A,B)$. This quantity can be computed by defining the joint probability distribution $P(A,B)$ for the two random variables $a,b$ belonging, respectively, to the $\sigma-$algebra of the subsystem $A$ and $B$, that we simply indicate with $a\in A$ and $b\in B$. \\ By defining the Shannon entropies $S_{Sh}$ as 
\begin{equation}
    \begin{split}
    S_{Sh}(A,B) &= -\sum_{a \in A, b \in B}p(a,b)\log(p(a,b))\\
    S_{Sh}(A) &= -\sum_{a \in A} p(a)\log(p(a))\\
    \end{split}
    \label{eq:shannon}
\end{equation}
with 
\begin{equation}
    \label{eq:marginalize}
    p(a) = \sum_{b \in B} p(a,b)
\end{equation}
we obtain the Shannon mutual information as
\begin{equation}
    I_{Sh}(A,B) = S_{Sh}(A) + S_{Sh}(B) - S_{Sh}(A,B)
    \label{eq:cmi}
\end{equation}
of which the von Neumann mutual information is a generalization. 

To obtain a classical distribution $P$ from a density matrix $\rho$, describing a quantum system, we can apply an orthogonal measurement $\mathcal{M}$ channel as follows
\begin{equation}
    \label{eq:measure}
    \begin{split}
    & p(i) = \bra{i}\rho \ket{i}\\ 
    &\mathcal{M}(\rho)=\sum_{i}P(i)\ket{i}\bra{i}
    \end{split}
\end{equation}
where the index $i$ runs over an orthogonal basis of $\mathcal{H}$, which in our case is the set of binary strings of length $n$.

For all practical purposes, retrieving the classical probability distribution defined by $\rho$ means considering the diagonal of $\rho$, or equally, using the probability distribution defined by the \acrshort{sd} in the wavefunction. The reduction to a space of interest of \eqref{eq:marginalize} is then a classical marginalization.

As detailed in appendix~\ref{app:A}, the two forms of mutual information introduced above are related by the following relationship:

\begin{equation}
    \label{eq:greater_inside_text}
    I_{vN}(A,B)\geq I_{Sh}(A,B)
\end{equation}

By using the two definitions of entropy for \eqref{eq:entropy} and \eqref{eq:shannon} we see that equality in \eqref{eq:greater_inside_text} holds if the state $\rho$ is already classical, i.e. it does not have off-diagonal elements.
Furthermore, we remark that in the established relations there was no mention to any particular subset of qubits, as these equations are valid $\forall  \mathscr{H}_{A},\mathscr{H}_{B} \subset \mathscr{H}_N$.

\subsection{Natural Orbitals}
\label{subsec:no}

In the determinant-based approaches for the calculation of the electronic molecular wavefunction, the choice of the 
one-electron basis set is a key aspect. Indeed, a transformation of the MO basis leads to the modification of the \acrshort{sd} ($\ket{\mathbf{i}}_{\mathcal{F}}$) coefficients in the expansion of the wavefunction eq.~\eqref{eq:ciwf}, even considering methods that are invariant under a unitary transformation of the MO basis.
\begin{equation}
   \ket{\psi}_{\mathcal{F}} = \sum_{\mathbf{i}=\{0,1\}^N} c_{\mathbf{i}} \ket{\mathbf{i}}_{\mathcal{F}} 
   \label{eq:ciwf}
\end{equation}
Overall, the nature of the \acrshort{mo} can have a marked effect on the numerical efficiency of the algorithm used to compute the wavefunction, such as the speed of convergence of the iterative Davidson procedure in \acrshort{ci} methods~\cite{David72}.

For molecular systems qualitatively well described by a single \acrshort{sd}, one can be tempted to base post-HF expansions on the \acrfull{hfco} basis set, which are the eigenvectors of the Fock matrix. By contrast, the \acrshort{hfco} have shown to be a poor one-electron basis, often exhibiting many non-negligible coefficients in \eqref{eq:ciwf} of similar magnitude in absolute value, indicating that a large number of \acrshort{sd}s has to be considered to approach 100\% of the wavefunction.

Since the beginning of quantum chemistry, an important effort has been devoted to defining a more effective orbital basis. Without climbing to be exhaustive, we cite here the Improved Virtual Orbital \cite{Hunt1969}, the Modified Virtual Orbital \cite{Bausc08}, and the Internally Consistent SCF \acrshort{mo} \cite{David72}.

Within the quest of an optimal one-electron basis set, a peculiar role is played by the \acrfull{no}.
The \acrshort{no} associated to a certain quantum state are the eigenvectors of the corresponding one-body reduced density matrix (eq.~\eqref{eq:1brdm}). As pointed out by Davidson \cite{David72} the virtual \acrshort{no} (those with an occupation number markedly lower than two) with the larger occupation numbers have the same extension as the occupied orbitals.
It has been conjectured  by L\"odwin in 1956 \cite{lowdin} that the use of the \acrshort{no} minimizes the number of Slater determinants required to describe a large percentage of the wavefunction in the CI expansion. This compactness leads to important advantages. First of all, the length of the CI expansion can be reduced without a significant decrease of its quality. This can be done by removing \acrshort{sd}s from the sum or by considering in the construction of the \acrshort{sd}s only the \acrshort{no} with the largest occupation numbers. Moreover, even while keeping unchanged the length of the CI expansion, the calculation of the wavefunction (irrespective of its nature, CI, PT, or CC) is expected to be more effective. It is worth noticing here, that also the analysis is simplified when the wavefunction is more compact (for instance opening the way to the application of the effective or intermediate Hamiltonian formalism \cite{Bloch58, Malri85}).

The construction of a CI wavefunction in its own natural orbital basis is not a trivial task, since the \acrshort{no} depend on the wavefunction which in turn is defined by the \acrshort{no}. In such a situation one can resort to an iterative procedure (\acrfull{ino})~\cite{Jafri1977} which can be expected to converge to \acrshort{mo} in which the wavefunction expresses a diagonal one-body reduced density matrix. Reaching self-consistency using this iterative method is rather expensive and in most case the procedure is stopped when a convergence criterion is reached.

On a side note, this iterative application is not always directly applicable, as in some cases the actual implementation of a method assumes that a specific \acrshort{mo} basis is used, such as \acrshort{hfco} in \acrshort{mp2}. For the sake of simplicity, for the following analysis, we only use \acrshort{ci} methods, which in the most popular implementation allow the use of any one-electron basis as input. It is worth noticing, however, that some adaptations enable other methods to satisfy this requirement (e.g. \acrshort{mp2}~\cite{finley00}).

\subsection{Two-qubit and one-qubit density matrices}
\label{subsec:1-2dm}

\begin{table*}[ht]
\small
\setlength\tabcolsep{1.8pt}
\centering
\newcommand{\minitab}[2][l]{\begin{tabular}#1 #2\end{tabular}}
\begin{tabular}{|c | c | c | c| c | c | c |}
\hline 
Mol. &  Geometry (\AA{})/source &  Basis-set & \#Qubits & \textit{ab initio}-method & $\chi_{max, HFCO}$ & $\chi_{max, NO}$\\ 
\hline
\hline
H$_2$O   & \minitab[c]{O 0.0 0.0 0.0 \\ H 0.757 0.586 0.0 \\ H -0.757 0.586 0.0}  & \minitab[c]{aug-ccpvtz \\ ccpvdz}  & \minitab[c]{184 \\ 48} & \minitab[c]{ \acrshort{cisd}\\ FCI} & \minitab[c]{$1.2\cdot10^{5}$\\$4.5\cdot10^{8}$} & \minitab[c]{$1.6\cdot10^{5}$\\$\cdot10^{5}$}\\
\hline
C$_2$   &  \minitab[c]{ 1.243 \\ 2.98\cite{Bytautas2015}} & aug-ccpvtz  & 184 & \acrshort{cisd} & \minitab[c]{$1.7\cdot10^{5}$\\$1.6\cdot10^{5}$} & \minitab[c]{$1.4\cdot10^{5}$\\$1.6\cdot10^{5}$ }\\
\hline
Be$_2$ &  4.6~\cite{khatib2014} & ccpv6z & 560 &\minitab[c]{ \acrshort{mrci}-CAS(4,8) \\ \acrshort{cisd}}& \minitab[c]{$2.0\cdot10^{5}$\\$3.2\cdot10^{5}$} & \minitab[c]{$7.7\cdot10^{4}$\\$3.5\cdot10^{5}$}\\
\hline
$(C_6H_6)_2$ &  \cite{Pitoňák2008} (s configuration) & ccpvdz & 456 & \acrshort{cisd} & $8.3\cdot10^{7}$ & $6.6\cdot10^{7}$ \\
\hline
\end{tabular}
\caption{Summary of tested molecular systems together with the computational details. The parameter $\chi_{max}$ is the number of \acrshort{sd} in the variational space for the solving method, out of $\chi_{max}$ \acrshort{sd}, we always took $min(\chi_{max}, 10^6)$ \acrshort{sd}. }

\label{tab: systems}
\end{table*}

It can be interesting to compare the approach here adopted with the formulation reported in the already mentioned works of Rissler \textit{et al}~\cite{Rissler2006} and Boguslawski~\cite{Boguslawski2015}, offering in this manner a direct comparison of mathematical relations with a purely computational approach following \eqref{eq:partial_trace}. More specifically, using these two past works we can establish some relations on the traced density matrices to one or two qubits. Let us first consider the two-qubit density matrices $\rho_{ij}$ for which there can be two separate cases, $i,j$ identify the occupation (through the Jordan-Wigner mapping\cite{Jordan1928}) of either two spin orbitals with opposite spin, marked by $i_{\uparrow}, j_{\downarrow}$, or the two spin orbitals with the same spin $i_{\uparrow}, j_{\uparrow}$, where in this notation the arrow is only used to point out the spin of the linked spin orbital, while $i,j$ can still vary from $0$ to $N-1$. For the former case, we begin with Ref.~\cite{Rissler2006}, where it is shown that such a density matrix is bound to be diagonal, so that it is of the form: 

\begin{equation}
    \label{eq:opposite_spin_diagonality}
    \rho_{i_{\uparrow}, j_{\downarrow}} = 
    \begin{pmatrix}
     \rho_{\ket{00}\bra{00}} & 0 & 0 & 0\\
    0 & \rho_{\ket{01}\bra{01}} &  0 & 0\\
    0 & 0 & \rho_{\ket{10}\bra{10}} & 0\\
    0 & 0 & 0 &  \rho_{\ket{11}\bra{11}}
    \end{pmatrix}
\end{equation}

This for us means exactly classical correlations since, as we already mentioned, eq~\eqref{eq:qmi} and eq~\eqref{eq:cmi} coincide for a diagonal $\rho$. 
This implies that the density matrix of whichever single qubit is also diagonal, as illustrated by \eqref{eq:single_diagonality}.

\begin{equation}
    \label{eq:single_diagonality}
    \rho_{i} = tr(\rho_{i,j})_j=
    \begin{pmatrix}
     \rho_{\ket{0}\bra{0}} & 0 \\
    0 & \rho_{\ket{1}\bra{1}}
    \end{pmatrix}
\end{equation}

It can then also be concluded that the entropy of a single qubit $i$ necessarily obeys $S_{vN}(i)=S_{Sh}(i)$.\\
At this point, from the definition of mutual information between qubit $i,j$ of eq.~\eqref{eq:qmi}, one has:
\begin{equation}
    I_{vN}(i,j) = S_{Sh}(i) + S_{Sh}(j) - S_{vN}(i,j)
\end{equation}
Furthermore, since $\rho_{i_{\uparrow}, j_{\downarrow}}$ is also diagonal as shown by eq.~\eqref{eq:opposite_spin_diagonality}, it also has purely classical correlations:

\begin{equation}
    I_{vN}(i,j) = S_{Sh}(i) + S_{Sh}(j) - S_{Sh}(i,j) = I_{Sh}(i,j)
\end{equation}
We have now remaining only the case of spin-orbitals with the same spin $i_{\uparrow}, j_{\uparrow}$, for which it is natural to ask if there are general diagonality conditions also for this last case. To this aim, starting from table~1 of~ref~\cite{Boguslawski2015, erratum} we can write:

\begin{equation}
    \label{eq:diagonality}
    \rho_{i_{\uparrow}, j_{\uparrow}} = 
    \begin{pmatrix}
     \rho_{\ket{00}\bra{00}} & 0 & 0 & 0\\
    0 &  \rho_{\ket{01}\bra{01}} &  \rho_{\ket{01}\bra{10}} & 0\\
    0 &  \rho_{\ket{10}\bra{01}} &  \rho_{\ket{10}\bra{10}} & 0\\
    0 & 0 & 0 &  \rho_{\ket{11}\bra{11}}
    \end{pmatrix}
\end{equation}

In this expression, components are defined in terms of multi-body density matrices. In particular, the \textit{off-diagonal} term $\rho_{\ket{01}\bra{10}}\equiv \rho_{\ket{10}\bra{01}}$ for $\rho_{i_{\uparrow}, j_{\uparrow}}$ is computed as:

\begin{equation}
    \label{eq:offdiagterm}
    \rho_{\ket{01}\bra{10}} = \prescript{}{\mathcal{F}}{\bra{\psi}}(a_{j}^{\dag}a_{i} -2a_{\bar{j}}^{\dag}a_{j}^{\dag}a_{i}a_{\bar{j}} )\ket{\psi}_{\mathcal{F}}
\end{equation}
where the barred index $\bar{j}$ is used to indicate spin-down orbitals indices and the expectation value takes place in the Fock space as in \eqref{eq:1brdm}\footnote{Technically, both the ladder operators and $\ket{\psi}$ are assumed to lye in the separable qubit space by~\cite{Rissler2006}, however this expectation is invariant under the fermionic-to-qubit mapping isometry.}.\\
In this expression the largest contribution is generally expected to come from the first term, $a_{j}^{\dag}a_{i}$, representing a looser condition than the second term for nullity of the expectation value between \acrshort{sd}s. For this reason, it is expected that decreasing (setting to zero) its value will likely bring to a more classical correlation pattern by bringing $\rho_{i,j}$ in a diagonal shape, and this is precisely the role of the \acrshort{no}.

\subsection{Computational Details}
\label{subsec:computationaldetails}

The molecules considered in this work were chosen to encompass systems with different kinds of chemical interactions and electronic correlation, such as vdW interactions, single and multiple bonds as reported in table \ref{tab: systems}. We used different quantum chemistry methods, depending on which is best suited to tackle each molecule.

All systems were first treated using a specific post \acrshort{hf} method in the restricted-\acrshort{hf} regime. In the \acrshort{mrci}SD method the wavefunction
is obtained diagonalizing the Hamiltonian on the space of all possible singles and doubles excitations out of the most important \acrshort{sd} of the CAS(\textit{e},N) solution.~\footnote{Actually, the \acrshort{fci} in CAS is made at configuration state function (CSF) level with a selection threshold for inclusion in MRCI of $\tau=10^{-5}$.}
As for the CASSCF part of this algorithm, it was only fully implemented for the simulation aiming at the natural orbitals, leaving aside orbital optimization (CASCI) for the simulation labeled by the \acrshort{hfco}

We use correlation consistent basis sets~\cite{Dunning1989}, using the aug-ccpvtz~\cite{Kendall1992} for H$_2$O and C$_2$, the ccpv6z basis for Be$_2$~\cite{khatib2014} ~\cite{Peterson1994, Wilson1996, Prascher2011} and ccpvdz~\cite{Pitoňák2008} basis set for $(C_6H_6)_2$. All basis set descriptions were taken from \url{basissetexchange.org}~\cite{Pritchard2019}.

We used different packages and codes for the simulations and data analysis, including the Pyscf~\cite{pyscf} python library, the Orca Package~\cite{Neese2018} for \acrshort{mrci} calculations and the SparQ~\cite{materia2024c} code mainly for the implementation of the trace~\eqref{eq:partial_trace} which has been implemented in-house in C++ to guarantee easy portability to parallelizing platforms such as OpenMP~\cite{Dagum1998}.

For all the systems we report the quantum and classical entropy and pair-wise mutual information for the spin-orbitals.  We undertake this analysis for two different sets of orbitals for each system, the \acrshort{hfco} and the \acrshort{no} (computed in the iterative scheme).
The analysis was carried out using the first $\chi$ \acrshort{sd} with the largest coefficient (in absolute value) in the expansion \eqref{eq:ciwf}. The chosen value of $\chi$ was however large ($10^6$), and in most cases includes all the \acrshort{sd} of the configurational space of the wavefunction. In Table~\ref{tab: systems} it is also reported the $\chi_{max}$ value, representing all possible \acrshort{sd} of non-zero coefficient.

\begin{figure*}
\vspace*{-1in}
    a)
\includegraphics[width=.95\textwidth]{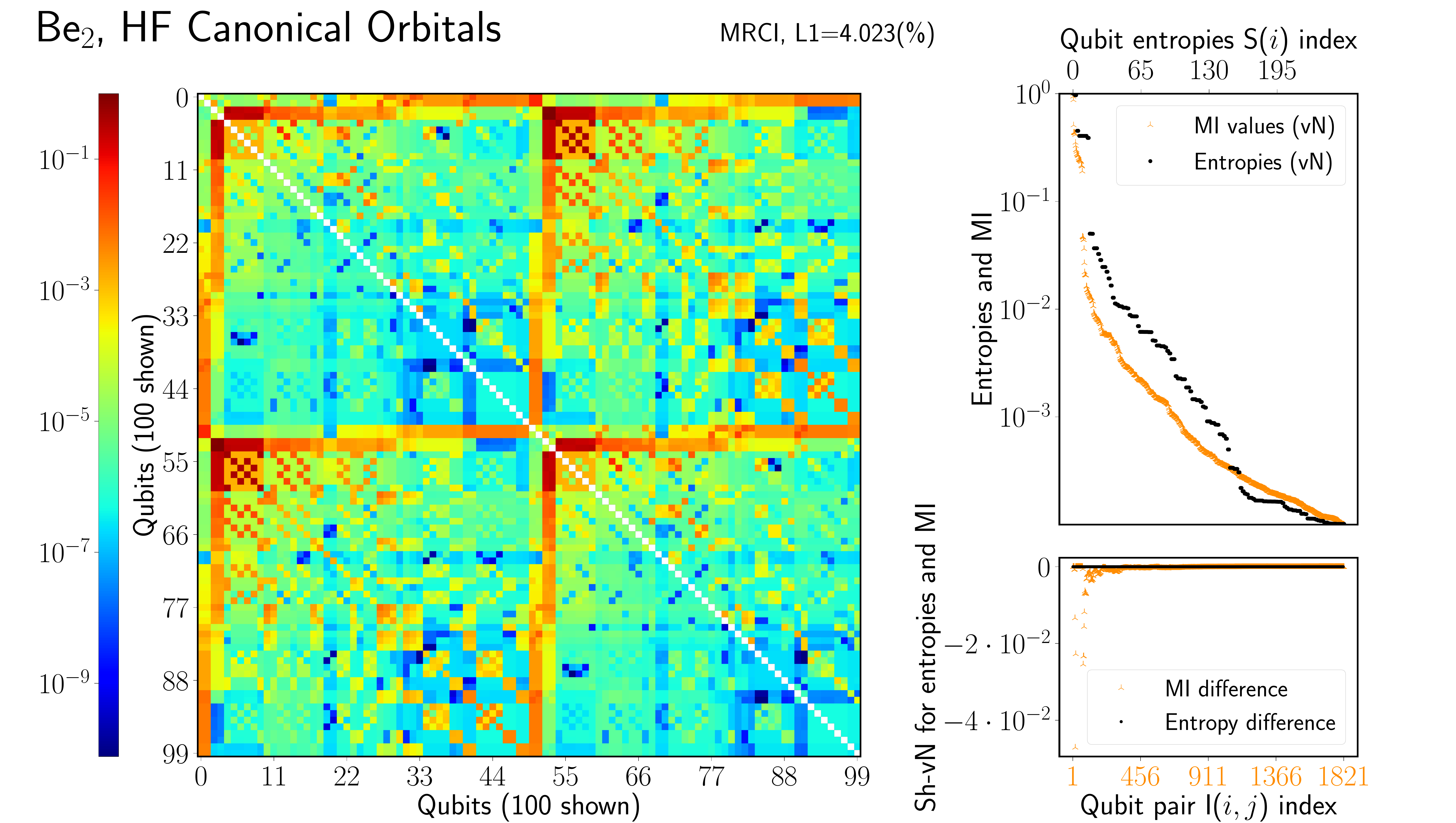}\hfill
        \\[\smallskipamount]

    \vspace{0.3cm}
    b)
\includegraphics[width=.95\textwidth]{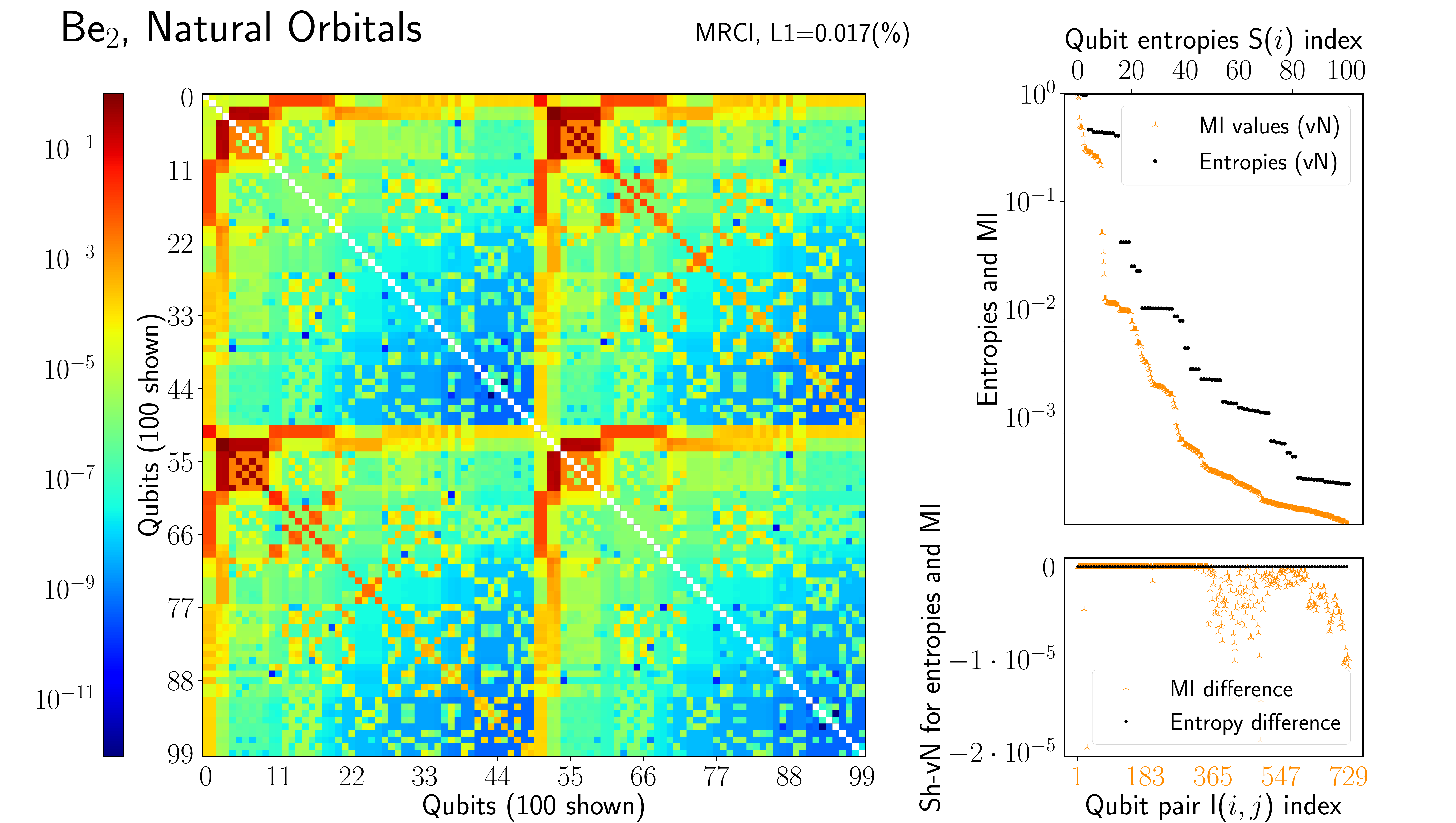}\hfill
    \caption{Classical and quantum orbital correlation for the MRCI wavefunction of Be$_2$ molecule expressed in Hartree-Fock Canonical Orbitals (top panel) and Natura Orbitals (bottom panel). \textit{Left}: Heatmap of the von Neumann mutual information $I_{vN}(i,j)$ values in the upper triangle of the matrix and of $I_{Sh}(i,j)$ in the lower triangle. To improve visualization we only displayed the 100 qubits/spin-orbitals with the largest entropy. \textit{Right top}: Plot of sorted elements $sort(I_{vN})$ of the $I_{vN}(i,j)$ matrix  (red empty circles) and sorted single qubit entropies $sort({S(i,j)})$(black filled circles); only values above a certain treshold are diplayed. \textit{Right bottom}: Difference of $sort(I_{Sh}) - sort(I)_{vN}$(orange) and corresponding differences between sorted single qubit entropies.}
    \label{fig:Be2}
\end{figure*}

\begin{figure*}
\vspace*{-1in}

    a)
\includegraphics[width=.95\textwidth]{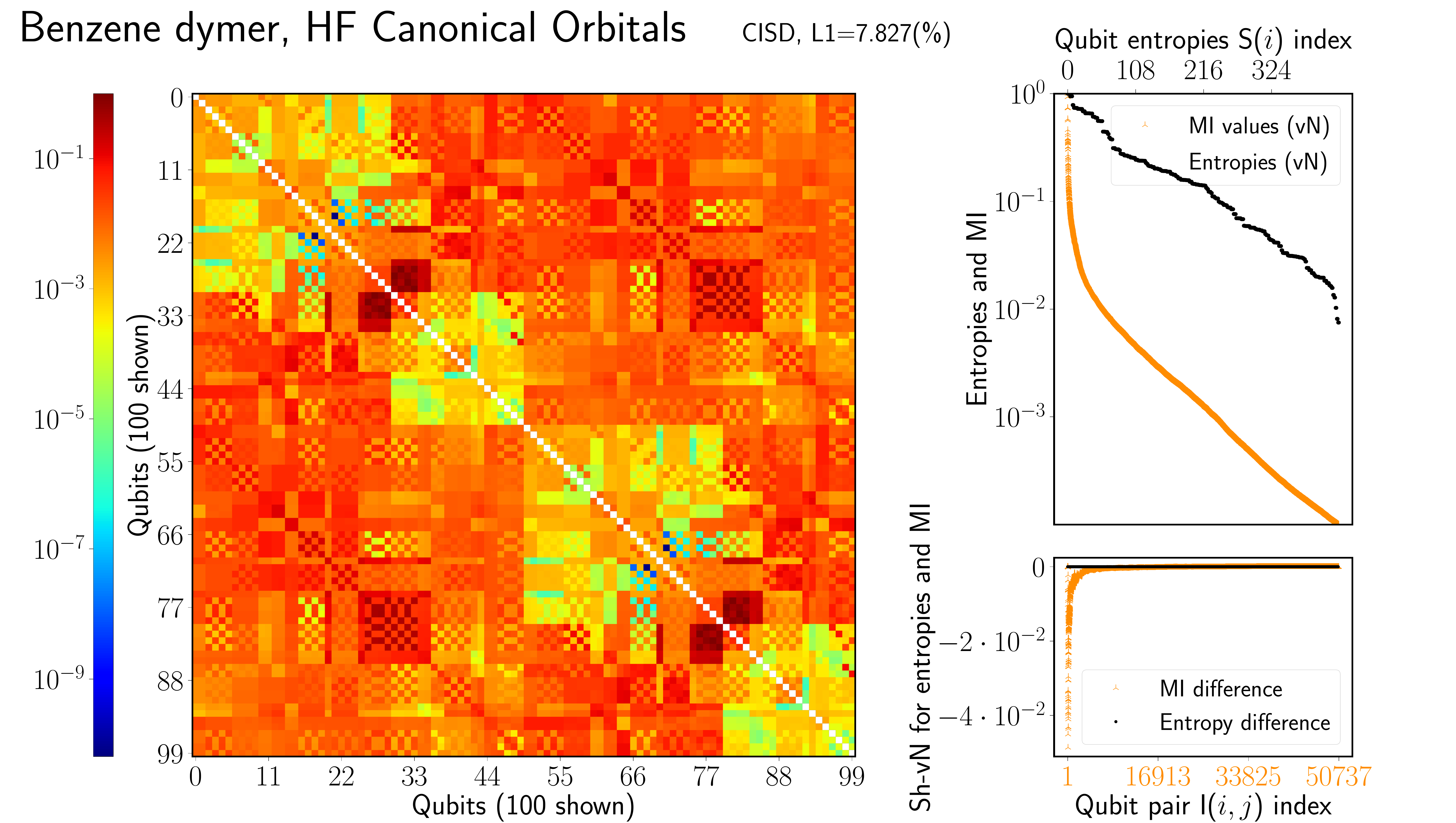}\hfill
    \\[\smallskipamount]

    \vspace{0.3cm}
    
    b)
\includegraphics[width=.95\textwidth]{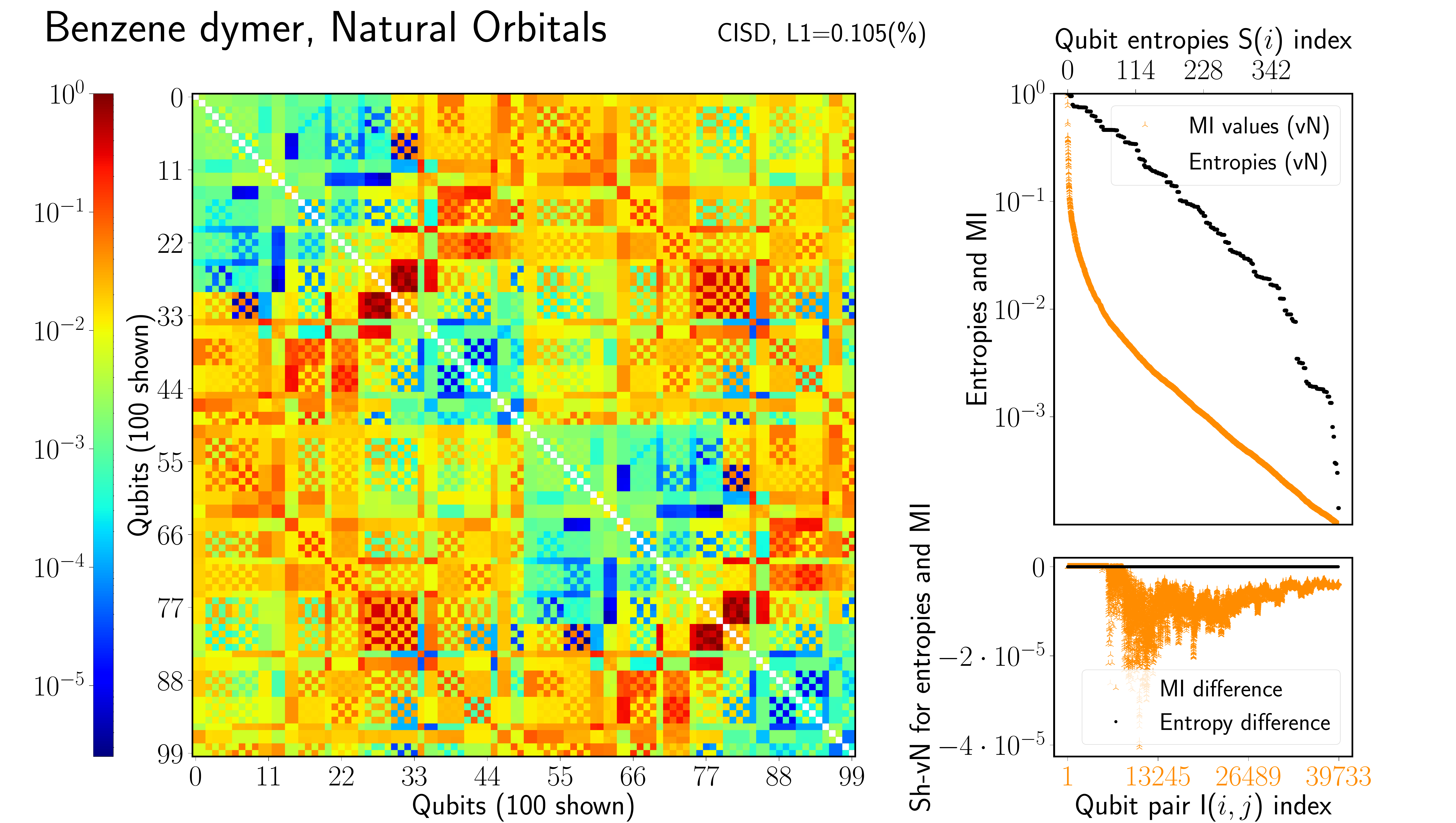}\hfill

    \caption{The two images report the result with the \acrshort{hfco} and the \acrshort{no} one-electron basis sets for the $(C_6H_6)_2$ system. \textit{Left}: Representation of the von Neumann mutual information $I_{vN}(i,j)$ values in the upper triangle of the matrix and of $I_{Sh}(i,j)$ in the lower triangle. $i,j$ are limited to the 100 qubits with the largest entropy. \textit{Right top}: Descending trend of $I_{vN}(i,j)$ matrix $\forall i,j$ (dark red) and same trend for the single qubit entropy $S_{vN}(i)$(black), both quantities are shown up to a $10^{-4}$ threshold. \textit{Right bottom}: Difference of $sort(I_{Sh}) - sort(I)$(orange) $\forall i,j$ and same comparison for all the single qubit entropies. }
    \label{fig:benzene}
\end{figure*}


\section{Results and Discussion}
\label{sec:results}

For our analysis of the quantum and classical information properties of the quantum chemistry's wavefunctions we have chosen molecular systems with different kinds of bonds and interactions, to investigate whether our findings might have a general relevance. 

We carried out a separate analysis for each system, delving into a comparison of the relevant quantities for our analysis. System description and computational details are summarized in table \ref{tab: systems}.
We begin to illustrate our results by focusing on figures ~\ref{fig:Be2} and ~\ref{fig:benzene} that contain two panels, the upper one referred to MRCI calculations on Be$_2$ at equilibrium distance using the Hartree-Fock Canonical Orbitals, the bottom one referred to the same calculation using Natural Orbitals. Within each panel, on the left is represented a heatmap plot of the quantum and classical mutual information matrix, showing for each qubit/spin-orbital index $i,j$ the values of the matrices, with the indices running over the 100 qubits with the largest entropy, of which the first 50 are linked to (via Jordan Wigner mapping) spin-up orbitals and the second half to spin-down orbitals. 
Exploiting the symmetry of the Mutual Information matrix we represented on the same heatmap both classical and quantum matrices. On the upper (top-right) triangular matrix, we have the von Neumann mutual information $I_{vN}(i,j)$ defined in~\eqref{eq:qmi}, whereas the lower (bottom-left) triangular matrix reports the Shannon Mutual Information $I_{Sh}(i,j)$ defined in~\eqref{eq:cmi}. 
This representation allows one to visually compare the differences between the two classical and quantum mutual information. 
A more detailed analysis is shown in the plots on the right side of the panels. Here, the upper part of the figure shows the diminishing sorted array $sort(I_{vN,\rho})(i,j)$ and single qubit entropies $sort(S_{vN,\rho})(i)$ for each $(i,j)$ pair and $i$, showing only values up to $10^{-4}$. 
Finally, in the lower plot of the panel, we represent the quantity $sort(I_{Sh,\rho}) - sort(I_{vN,\rho})$ with the sorting running over indices $i,j$, thus the subtraction between the sorted array of $I_{Sh,\rho}$ and the curve on the top plot. A similar analysis is also reported for entropy, thus showing the difference of $sort(S_{Sh,\rho}) - sort(S_{vN,\rho})$ with the sorting running over the qubit index $i$.

Figure 2 shows the same data for the benzene dimer example. Both series of data in Figure~\ref{fig:Be2} and ~\ref{fig:benzene} showdifferences of the structure of the von Neumann mutual information between the \acrshort{hfco} and the \acrshort{no}, namely the differences between the upper-right triangular part of the heatmap and the lower-left triangular part. It is already clear visually that the differences are significantly attenuated in the lower panels, where the wavefunctions are expressed in terms of Natural Orbitals.
This is in line to what we already discussed in section~\ref{subsec:no}, since the \acrshort{no} are supposed to provide, in general, a basis set in which the wavefunction is represente in a more compact structure.

The difference between the von-Neumann  and Shannon quantities is  larger in the \acrshort{hfco} basis than in their \acrshort{no} counterparts, as visually perceptible in Figure res~\ref{fig:Be2}(a) and~\ref{fig:benzene}(a). Numerically this effect can be estimated numerically using the following expression:
\begin{equation}
    \label{eq:diff}
    L1_\rho=\frac{\sum_{i,j>i}^{(i,j)=N} I_{vN}(i,j)-I_{Sh}(i,j)}{\sum_{i,j>i}^{(i,j)=N} I_{vN}(i,j)}*100
\end{equation}
The values of the $L1_\rho$ parameter for the different molecules using both the \acrshort{hfco} and the \acrshort{no} are reported in Tab.~\ref{tab: systems_results}.

\begin{table*}[ht]
\small
\setlength\tabcolsep{1.8pt}
\centering
\newcommand{\minitab}[2][l]{\begin{tabular}#1 #2\end{tabular}}
\begin{tabular}{| c | c | c | c | c | c | c | c | c |}
\hline 
Mol. &  \minitab[c]{ Div. \\ summary} & $E_{HFCO}(E_h)$& $E_{NO}(E_h)$& \minitab[c]{ $L1_\rho$ \\ HFCO} & \minitab[c]{ $L1_\rho$ \\ NO} & $\gamma_{HF}$& $\gamma_{NO}$\\ 
\hline
\hline
H$_2$O  & \minitab[c]{ aug-ccpvtz\\ccpvdz (\acrshort{fci})} & \minitab[c]{-76.337\\ -76.244} & \minitab[c]{-76.336\\-76.244} &\minitab[c]{20.942 \\ 5.0508} & \minitab[c]{0.058\\0.060}  & \minitab[c]{ 0.132 \\ 0.260 } & \minitab[c]{$\sim10^{-8}$\\$\sim10^{-9}$}\\
\hline
C$_2$  & \minitab[c]{ 1.243\AA\\2.98\AA } & \minitab[c]{-75.737\\-75.433} & \minitab[c]{-75.733\\-75.428} &\minitab[c]{ 16.753 \\ 10.857} & \minitab[c]{0.011\\0.006} & \minitab[c]{ 0.150 \\ 0.147} & \minitab[c]{$\sim10^{-6}$ \\$\sim10^{-5}$}\\
\hline
Be$_2$  & \minitab[c]{ \acrshort{mrci} \\ \acrshort{cisd}} & \minitab[c]{-29.307\\-29.293} & \minitab[c]{-29.306\\-29.293} & \minitab[c]{ 4.023\\ 18.587} & \minitab[c]{0.017\\0.001} & \minitab[c]{//\\0.127} & \minitab[c]{//\\$\sim10^{-5}$}\\
\hline
(C$_6$H$_6$)$_2$ &  & -462.627 & -462.627 & 7.827 & 0.105 & 0.023 & $\sim 10^{-4}$ \\
\hline
\end{tabular}
\caption{Summary of results for tested systems, the second column reports the division 
as shown in tab.~\ref{tab: systems}. $L1_\rho$ in eq~\eqref{eq:diff} has been introduced as percentage difference in correlations between \eqref{eq:qmi} and \eqref{eq:cmi} (see eq.~\eqref{eq:diff}).The $\gamma$ parameter is introduced in eq.~\eqref{eq:off}.}
\label{tab: systems_results}
\end{table*}

The $L1_\rho$ difference (in percentage) is a quantity always greater or equal to zero because ofthe inequality~\eqref{eq:greater_inside_text}.  We reported $L1_\rho$ both in the caption of Figures~\ref{fig:Be2} and~\ref{fig:benzene} as percentage and in table~\ref{tab: systems} as absolute values
The reported computed data clearly shows that Natural Orbitals one-particle basis set gives a description of the correlation which is essentially classical, drastically reducing the $L1_\rho$ difference between the von Neumann and the Shannon mutual information.

It is worth noticing that all the above analyses are based on approximate wavefunctions, and, moreover, that the Natural Orbitals are also approximate because of two different reasons.
Natural orbitals are indeed obtained by an iterative procedure, namely, they are Iterative Natural Orbitals. The approximations comes therefore from two distinct sources: the first is the order of the iteration, the second is due to the fact that the ground state method is also approximated by the post Hartree-Fock method used in the specific case.
The convergence as a function of the number of iterations can be followed by monitoring a parameter that quantifies the diagonal shape of the density matrix at each iteration :

\begin{equation}
    \label{eq:off}
    \gamma = \frac{1}{N_{e}} \sum_{i,j, i\neq j}^{N}|\rho_{i,j}|
\end{equation}
with $N_{e}$ being the number of electrons (and therefore also the trace of the one-body reduced density matrix).

\begin{figure}[h!]
    \centering
    \includegraphics[width=0.5\textwidth]{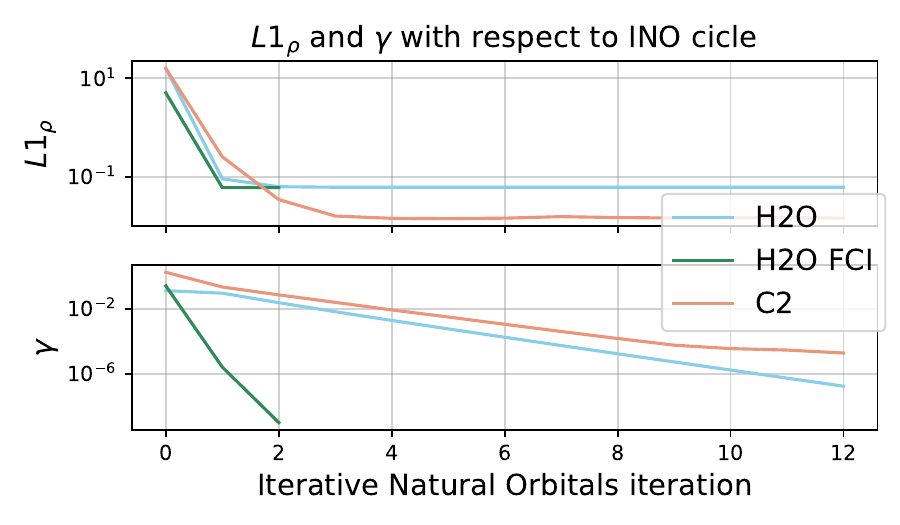}
    \caption{Improvement of iterative natural orbital through different iterations via $\gamma$ and the effect on the resulting L1.}
    \label{fig:noconvergence}
\end{figure}


We report in Figure~\ref{fig:noconvergence} how this quantity changes along the iterative procedure. It is clear that the \acrshort{no} procedure properly converges. It is worth noticing that during the process the energy can also increase since the procedure is not variational, an example is shown in tab~\ref{tab: systems_results} by the molecules H$_2$O, C$_2$ and Be$_2$, however, moving from the \acrshort{hfco} to the \acrshort{ino} basis the energy increase is limited to a few hundredths of a Hartree, at most. 
On the other hand, we noted that stopping the iterations at the first step rarely increases the energy, but $\gamma$ can still display quite a large value, as clearly shown in Figure \ref{fig:noconvergence}. 

Since we are interested in reducing the off-diagonal terms of the density matrix, which are sensitive to the accuracy of the \acrshort{no}, to guarantee to use properly converged natural orbitals, we carried on the iterative procedure until either $\gamma$ was lower than $\sim10^{-8}$, or a maximum of about $\sim10$ iterations were reached. 
The value of $\gamma$ at the last iteration is also reported in table~\ref{tab: systems_results}, from which one can highlight the marked difference of the $\gamma$ value of the \acrshort{hfco} and \acrshort{ino}.

The convergence of $\gamma$ and the corresponding convergence of the $L1_\rho$ values are reported in figure~\ref{fig:noconvergence}.

In the previous section, we have reported a solid evidence for the classical behavior of the orbital-wise correlation when the wavefunction is expressed in the basis of natural orbitals. For example, the heatmap plots of figures~\ref{fig:Be2} and~\ref{fig:benzene} are a clear visual representation of the effect of the one-electron basis set on the von Neumann mutual information and its relation with the correspondin classical mutual information based on the Shannon entropy. 
The $L1_\rho$ value, which is providing an estimate of the differences between classical and quantum correlation, is significantly smaller in Natural Orbitals with respect to its value in the Hartee-Fock basis (up to 100 times in some cases).
When sorted, the values of mutual information (red spots in the right panels in the figure) decrese much faster using NO than with \acrshort{hfco}. This result has been already pointed out by us recently~\cite{Ratini2023b} but here is tested on a different range of systems and larger molecular sizes.

From the computational point of view, the classical behavior of correlations in the \acrshort{no} in principles basis paves the way for a simplification of the computational resources required to attain a given precision in the simulation. 
Two important implications arise.
Firstly, these findings show, once more, the importance of the choice of orbital basis for \acrshort{ci}-based method.  In this regard, the use of \acrshort{no} in classically computed quantum chemistry is strongly suggested whenever possible since their key role in CI calculation has been longly recognized. However, we believe that their potential should not be limited to such methods. In Quantum Computing, we have shown how the use of \acrshort{no} can be beneficial for the efficiency of variational algorithms \cite{Materia2024, Ratini2023b}, and the analysis reported in this work confirms these considerations. Because these properties, it is clear that Natural Orbitals should always be preferred to properly analyze the wavefunction in terms of orbital-correlation. Indeed, other orbital choices would likely overestimate the quantum contribution to the mutual information.

The second implication delves into the foundation of quantum computing. The second implication delves into the foundations of quantum computing. If correlations in quantum chemistry are predominantly classical in nature, additional attention is warranted to the theoretical need for a quantum computer to solve the electronic structure problem.
In line with these concerns, other authors have expressed skepticism and reservations about the actual potential of quantum computing in quantum chemistry.~\cite{Lee2023}. 

Concerning the classical structure of the von Neumann mutual information when the wavefunction is built on \acrshort{no}, the results here reported are consistent with the discussion developed in section \ref{subsec:1-2dm} on the relationships described in \eqref{eq:offdiagterm} and based on chemist's multi-bodies reduced density matrices to determine the traced density matrices required to calculate the von Neumann mutual information. Not only we confirm what is described by eq.~\eqref{eq:offdiagterm}, 
 but also validate all the contributing factors that led to this conclusion.
 Specifically, the entropies of single qubits are characterized entirely by classical entropies ($S_{vN}=S_{Sh}$), as shown by top right of figures \ref{fig:Be2} and~\ref{fig:benzene}. Additionally, the mutual information between two qubits, when mapped to orbitals with opposite spins, also exhibits a purely classical characterization, independently on the spin orbital basis.
For the remaining combination of same-spin qubits, some residual quantum contribution to the mutual information is remaining, even if this is significantly reduced (up to 100 times) in the fully converged \acrshort{ino} basis. This residual, which is still present with fully converged \acrshort{ino}s, 
Considering one body reduced density matrix diagonal, the orginin of this  residual $L_1$ difference might be due to the higher order multi-body reduced density matrices, as pointed out by quation~\eqref{eq:offdiagterm}. 

\subsection{Conclusions}
\label{subsec:conclusions}


In summary, this work investigated whether the orbital-wise correlations in Quantum Chemistry are quantum or classical in nature, employing techniques from Quantum Information Theory. We conducted numerical experiments on moderate-sized systems, which vary in chemical nature and types of bonding. These systems were treated using high-level post-Hartree-Fock methods and involved up to approximately 550 spin-orbitals.
Our numerical tests, supported by mathematical reasoning, illustrate how Natural Orbitals significantly reduce the purely quantum von Neumann-only qubit-qubit (spin orbital-spin orbital) correlations, predominantly leaving only the classical contributions.
The computational and theoretical evidence obtained in this study lays the foundation for computational simplification in the processing of quantum information quantities in quantum chemistry. This provides additional mathematical justification for the more widespread use of Natural Orbitals and raises further questions about the actual computational complexity of the multi-body problem in quantum chemistry.

\subsection{Acknowledgments}
The authors acknowledge funding from the MoQS program, founded by the European Union’s Horizon 2020 research and innovation under the Marie Skłodowska-Curie grant agreement number 955479.
The authors acknowledge funding from Ministero dell’Istruzione dell’Università e della Ricerca (PON R \& I 2014-2020).
The authors also acknowledge funding from National Centre for HPC, Big Data and Quantum Computing - PNRR Project, funded by the European Union - Next Generation EU.\\
L.G. and C.A. acknowledge funding from the Ministero dell'Università e della Ricerca (MUR) under the Project PRIN 2022 number 2022W9W423 through the European Union Next Generation EU.

\newpage
\begin{appendices}
\section{}
    \label{app:A}
\setcounter{equation}{0}
\numberwithin{equation}{section}

We hereby show the relation $I_{vN}\geq I_{sh}$ reported in eq~\eqref{eq:greater_inside_text}.\\
Starting from eq.~\eqref{eq:measure} and given a \acrfull{tpcpm}\footnote{Complete positivity affirms positivity of the operator up to any space extension besides $A,B$.~\cite{Watrous2018}}, indicated with $\mathcal{O}$, i.e. a local operation applied on a subspace, the following can be proven~\cite{Watrous2018}:
\begin{theorem}
    \label{thm:monotonicity}
    Given a $\rho_{AB}\in \mathscr{H}_A \otimes \mathscr{H}_B$ and a \acrshort{tpcpm} $\mathcal{O}$ acting only on space B, then:
    \begin{equation}
        \label{eq:monotonicity}
            I_{\rho}(A,B)\geq I_{\mathcal{O}\left(\rho_{AB}\right)}(A,B)
    \end{equation}
\end{theorem}

This relation states the intuitive fact that a local operation only on space $B$ cannot increase the mutual information between $A$ and $B$.
In the case under analysis, a measurement can be considered as a \acrshort{tpcpm},  since it is both trace-preserving and it doesn't alter the positivity of the density matrix. \\
For a measurement applied on a single qubit, e.g. at qubit $0$ ($\mathcal{M}(\rho)_{0}$), eq.~\eqref{eq:monotonicity} then becomes 

\begin{equation}
    \label{eq:greater}
    I_{vN}(A,B)\geq I_{\mathcal{M}(\rho)_{0}}(A,B)
\end{equation}
Which repeated to all the qubits one at a time leads to:

\begin{equation}
    \label{eq:greater_1by1}
    \begin{split}
        &I_{\mathcal{M}(\rho)_{0}}(A,B) \dots \geq I_{\mathcal{M}(\rho)_{0,\dots i}}(A,B) \\
        &\geq I_{\mathcal{M}(\rho)}(A,B) = I_{Sh}(A,B) 
    \end{split}
\end{equation}
where $\mathcal{M}(\rho)$ is the measurement over all the qubits. The last equality of \eqref{eq:greater_1by1} is because a projective measurement $\mathcal{M}$ renders diagonal $\rho$ by discarding all the off-diagonal terms, but \eqref{eq:qmi} coincides with \eqref{eq:cmi} for diagonal matrices. 
This finally leads us to:

\begin{equation}
    \label{eq:greater_2}
    I_{vN}(A,B)\geq I_{Sh}(A,B)
\end{equation}

\end{appendices}

\newpage

\bibliographystyle{unsrt}

\clearpage

\printglossary[type=\acronymtype]
\printglossary

\end{document}